# Closed-Form Critical Conditions of Saddle-Node Bifurcations for Buck Converters

Chung-Chieh Fang

## Abstract

A general and exact critical condition of saddle-node bifurcation is derived in closed form for the buck converter. The critical condition is helpful for the converter designers to predict or prevent some jump instabilities or coexistence of multiple solutions associated with the saddle-node bifurcation. Some previously known critical conditions become special cases in this generalized framework. Given an arbitrary control scheme, a systematic procedure is proposed to derive the critical condition for that control scheme.

C.-C Fang is with Advanced Analog Technology, 2F, No. 17, Industry E. 2nd Rd., Hsinchu 300, Taiwan, Tel: +886-3-5633125 ext 3612, Email: fangcc3@yahoo.com

## I. INTRODUCTION

The buck converter, a type of DC-DC converter, is a nonlinear *hybrid* system which exhibits both continuous and discrete dynamic behavior. It is also a *piecewise* linear system [1], [2], [3] that the dynamics is linear in each time segment separated by the switching actions of the converter. Due to the switching actions involved in the operations, the converter is more accurately modeled as a sampled-data system [1], [2], [3]. Instability occurs when there exists a sampled-data pole outside the unit circle in the complex plane.

There are three ways that the sampled-data pole leaves the unit circle, thus causing three typical instabilities in DC-DC converters [4], [5], [6]. When the sampled-data pole leaves the unit circle through 1 in the complex plane, the instability is generally a saddle-node bifurcation (SNB) [5] (or pitchfork and transcritical bifurcations [6] which are not typically seen in DC-DC converters). The SNB generally involves coexistence of multiple solutions [7], or sudden disappearances or jumps of steady-state solutions [5]. When the pole leaves the unit circle through -1, the instability is a period-doubling bifurcation (PDB) [8], which generally involves *fast-scale* subharmonic oscillation. It is a phenomenon where the signal ripple has subharmonics. When the pole leaves through a point other than 1 or -1 on the unit circle, the instability is a Neimark-Sacker bifurcation (NSB), which generally involves a *slow-scale* quasi-periodic oscillation [3], [5]. This paper focuses on SNB.

Averaged models are generally applied to analyze DC-DC converters. It has been known that most averaged models cannot accurately predict the occurrence of PDB (subharmonic oscillation) [8]. By considering the sampling effects and increasing the system dimension, improved averaged models can predict the occurrence of PDB [9], [10]. However, even with positive gain or phase margins in the *improved averaged* model, instability may still occur [11]. Therefore, there is a strong need to obtain the *exact* critical condition of instability.

Typical control schemes commonly used in DC-DC converters are voltage mode control (VMC) and current mode control (CMC) [12]. Some SNB critical conditions for *particular* control schemes have been known. For example, in CMC with light loading, an SNB critical condition for the buck converter has been reported [13]. This paper tries to answer the following questions:

1) Is there a *general* closed-form critical condition that directly leads to these known conditions, and these known conditions become *special* cases in the generalized framework?
2) Some control schemes (such as VMC and CMC) may seem different, but do they share the same or similar *form* of critical condition?
3) Given an *arbitrary* control scheme, is there a *systematic* method to derive the critical condition for that control scheme?

The answers to all of these questions will be shown to be affirmative.

Like the describing function approach [14], harmonic balance [15], [16] is a tool to analyze a nonlinear system. Based on harmonic balance, an PDB critical condition for the buck converter is obtained in [3], [17], [18]. Following similar approach, in this paper, the *exact* SNB critical condition is derived.

Note that, here, the critical condition is expressed in the converter *parameter space*. For example, given a source voltage $v_s$ and load resistance $R$, an illustrative critical boundary in the parameter space $(v_s, R)$ is shown in Fig. 1. The critical conditions defines the SNB boundary in the parameter space to separate stable and unstable regions. When a converter parameter crosses the critical value, the stability (or instability) changes. Critical conditions in *closed-forms* greatly facilitate the converter design, because the *quantitative* effect of each *relevant* converter parameter can be clearly seen.

The remainder of the paper is organized as follows. In Sections II and III, the operation of the buck converter and harmonic balance analysis are presented. A general critical condition in terms of loop gain is derived. In Section IV, the critical conditions for typical loop gains are derived. In Sections V-VI, the proposed approach is systematically applied to various control schemes. Conclusions are collected in Section VII.

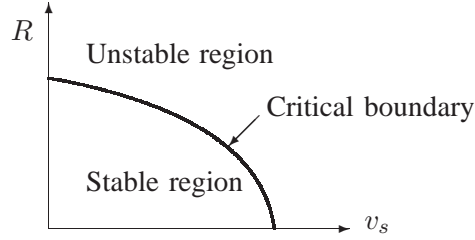

Figure 1. An illustrative critical boundary in the *parameter* space $(v_s, R)$.

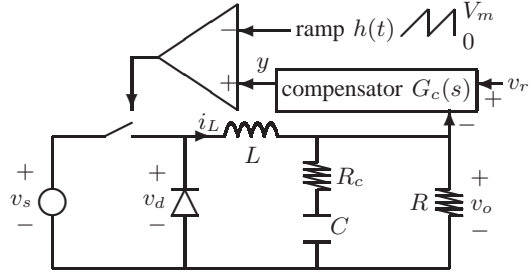

Figure 2. A VMC buck converter with a compensator $G_c(s)$.

## II. BRIEF REVIEW OF BUCK CONVERTER OPERATION

Consider a VMC buck converter shown in Fig. 2, where $v_s$ is the source voltage, $v_d$ is the voltage across the diode, $v_o$ is the output voltage, $v_r$ is the reference voltage, $y$ is the compensator output signal, and $G_c(s) = -y(s)/v_o(s)$ is the compensator transfer function. Denote the ramp amplitude as $V_m$, and the ramp slope as $m_a = \dot h(d)$. Denote the cycle period as $T$, the switching frequency as $f_s = 1/T$ and let $\omega_s := 2\pi f_s$.

Similarly, a CMC buck converter is shown in Fig. 3, where a control signal $i_c$ controls the (peak) inductor current $i_L$, and $y = i_c - i_L$ (equivalently, $G_c(s) = 1$). Note that in Fig. 3, the circuit is arranged in a way to have a similar output signal $y$ as in VMC.

The operation of the converter is as follows. Within a cycle period $T$, the dynamics is switched between two stages, $S_1$ and $S_2$. Switching occurs when the compensator output $y$ intersects with the ramp signal $h(t) := V_m(\frac{t}{T} \bmod T)$. One has $v_d = v_s$ in stage $S_1$ and $v_d = 0$ in stage $S_2$. The waveform of $v_d(t)$ is a square wave with a duty cycle $D$. The controlled buck converter is equivalent to a nonlinear switching model shown in Fig. 4. Note that $v_r$ adjusts the DC offset and does not affect the loop stability. It can be implicitly modeled to adjust the offset of $h(t)$. Let the loop gain be $T(s)$. From Fig. 4, $T(s) = v_s G(s)/V_m$, and the model in Fig. 4 can be normalized as shown in Fig. 5.

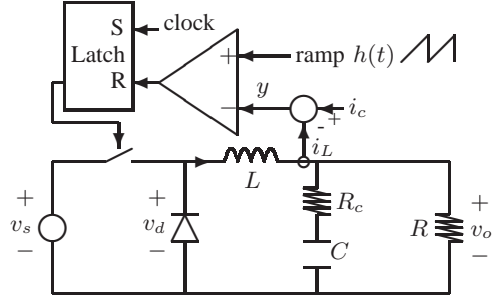

Figure 3. A CMC buck converter.

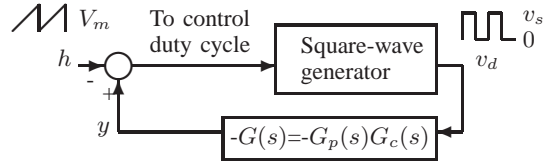

Figure 4. An equivalent nonlinear switching model.

## III. GENERAL CRITICAL CONDITION DERIVED FROM HARMONIC BALANCE ANALYSIS

The signal path in the control *loop* has two parts: from $y$ to $v_d$ through a *nonlinear* PWM modulator, and from $v_d$ to $y$ through a *linear* transfer function $G(s) := G_p(s)G_c(s)$, where $G_p(s)$ is the power stage transfer function and its representation depends on what signal is fed to the compensator.

In VMC, $v_o$ is fed to the compensator. Let $\rho = R/(R + R_c)$. For $R_c = 0$, $\rho = 1$. From [12, p. 470], the power stage $v_d$-to-$v_o$ transfer function is

$$G_p(s) = \frac{sR_cC + 1}{\frac{LCs^2}{\rho} + (\frac{L}{R} + R_cC)s + 1} \tag{1}$$

In CMC, $i_L$ is fed to the compensator (with $G_c(s) = 1$). From [12, p. 470], the power stage $v_d$-to-$i_L$ transfer function is

$$G_p(s) = \frac{\frac{Cs}{\rho} + \frac{1}{R}}{\frac{LCs^2}{\rho} + (\frac{L}{R} + R_cC)s + 1} \tag{2}$$

Let $x^0(t)$ be the $T$-periodic solution of the converter. Let $y^0(t)$ be the corresponding $T$-periodic compensator output signal. The intersection of $h(t)$ with $y^0(t)$ determines the duty cycle and hence the waveform of $v_d(t)$. By "balancing" the equation $y^0(t) = h(t)$ (written in Fourier series form) at the switching instants, conditions for existence of periodic solutions and SNB can be derived.

Let $d = DT$. In steady state as shown in Fig. 6, $v_d(t)$ is $T$-periodic and it can be represented by Fourier series (harmonics)

$$v_d(t) = v_s \sum_{n=-\infty}^{\infty} c_n e^{jn\omega_s t} \text{ where } c_n = \frac{1 - e^{-jn\omega_s d}}{j2n\pi} \tag{3}$$

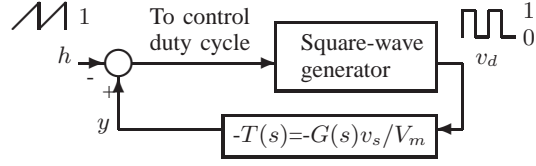

Figure 5.  An equivalent *normalized* nonlinear switching model.

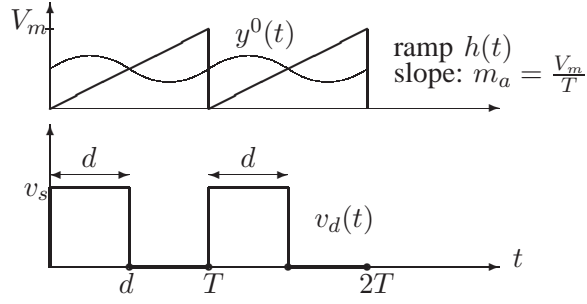

Figure 6.  Illustrative signals of $y^0(t)$, $h(t)$ and $v_d(t)$.

From Figs. Fig. 2 and 4, one has

$$y^0(t) = G_c(0)v_r - v_s \sum_{n=-\infty}^{\infty} c_n e^{jn\omega_s t} G(jn\omega_s) \tag{4}$$

Since $y^0(t)$ intersects with $h(t)$ at $t = d = DT$, one has $y^0(d) = h(d)$. Using (3), one has

$$G_c(0)v_r - \sum_{n=-\infty}^{\infty} \frac{v_s}{j2n\pi}(e^{jn\omega_s d} - 1)G(jn\omega_s) = h(d) \tag{5}$$

This is an equation of $d$. When SNB occurs, only a *single* solution exists, the curve of $y^0(d)$ is tangent to $h(d)$. The SNB is also called a tangent bifurcation [6]. One has $\dot{y}^0(d) = \dot{h}(d) = m_a$, which leads to

$$-\frac{2v_s}{T} \sum_{n=-\infty}^{\infty} e^{jn\omega_s d} G(jn\omega_s) = m_a \tag{6}$$

Note that, using (3) and the fact that $y^0(d) = h(d)$,

$$v_s = \frac{G_c(0)v_r - h(d)}{\sum_{n=-\infty}^{\infty} c_n e^{jn\omega_s d} G(jn\omega_s)} \tag{7}$$

which shows $v_s$ as a function of $d = DT$.

Let **Re** denote taking the real part of a complex number. Note that $\omega_s d = 2\pi D$, $e^{jn\omega_s d}G(jn\omega_s) + e^{-jn\omega_s d}G(-jn\omega_s) = 2\mathbf{Re}[e^{jn\omega_s d}G(jn\omega_s)]$, and $T(s) = G(s)v_s/V_m = G(s)v_s/m_a T$. Then, by simple algebra, (6) leads to the following theorem.

**Theorem:** *Consider a closed-loop buck converter shown in Fig. 4 or Fig. 5. The critical condition of SNB is*

$$-\frac{2v_s}{T}\mathbf{Re}[\sum_{n=1}^{\infty} e^{j2n\pi D}G(jn\omega_s)] - \frac{v_s G(0)}{T} = m_a \quad (8)$$

*For $V_m \neq 0$, (8) is expressed in terms of loop gain as*

$$-2\mathbf{Re}[\sum_{n=1}^{\infty} e^{j2n\pi D}T(jn\omega_s)] = T(0) + 1 \quad (9)$$

It will be shown later that these critical conditions can be expressed in a closed-form related to the hyperbolic function csch. Note that (9) is an expression of convenience. For $V_m = 0$, the loop gain $v_s G(s)/V_m$ would be infinite. In that case, the equivalent critical condition (8) is used. Since both (8) and (9) are exact critical conditions, they can be used as benchmarks to determine the accuracy of other critical conditions.

Generally the power stage has an order of at least two (associated with $L$ and $C$). If the compensator has an order of three, the loop gain $T(s)$ has an order of five. However, the compensator is generally designed to cancel some poles or zeros of the power stage, and the order of $T(s)$ is reduced. Note that the right side of (9) involves $T(0)$, but the left side of (9) involves only frequency higher than $\omega_s$. The shape of the Bode plot for the frequency smaller than $\omega_s$ is irrelevant. One can use a simplified high-frequency form of $T(s)$ for stability analysis. By decomposing the loop gain into *partial fractions*, many new *closed-form* critical conditions in terms of the converter parameters can be obtained.

**L-plot and S-plot.** Define an "F-transform" of $T(s)$ as

$$\mathcal{F}[T(s)] := -2\mathbf{Re}[\sum_{n=1}^{\infty} e^{j2n\pi D}T(jn\omega_s)] \quad (10)$$

Note that $\mathcal{F}[T(s)]$ is a function of many converter parameters. Here, this function is called an L-plot, denoted as $\mathcal{L}$. For example, let it be a function of $D$ and it becomes $\mathcal{L}(D)$. Then, the critical condition (9) becomes $\mathcal{L} := \mathcal{F}[T(s)] = T(0)+1$. The critical condition itself does not tell which side of the critical boundary will be the stable region. Generally for a converter, the region with $\mathcal{L} < T(0)+1$ is stable. Also, define an S-plot as $\mathcal{S} = m_a \mathcal{L}/(T(0)+1)$, which is also the left side of (6) or (8). Then, the critical condition (8) becomes $\mathcal{S} = m_a$, and the region with $\mathcal{S} < m_a$ is stable. The S-plot is a useful design tool because it shows the minimum ramp slope to stabilize the converter. The F-transforms of typical loop gain functions are presented next.

## IV. "F-TRANSFORMS" OF TYPICAL LOOP GAIN FUNCTIONS

The loop gain is generally designed to have sufficient gain and phase margins. For a stable converter, the phase of $T(j\omega)$ is less than $180°$ at the crossover frequency $\omega_c$. One can focus only those loop gains of first or second orders.

The loop gain can be further decomposed into a combination of partial fractions. Only partial fractions of first orders are considered. Similar analysis can be applied to partial fractions of second orders.

Let $\omega_p$ and $\omega_z$ be the pole and zero of $T(s)$. Let $p = \omega_p/\omega_s$ and $z = \omega_z/\omega_s$. The F-transform of the fraction $1/(1+\omega_p)$ will be the building block to derive the F-transforms of other loop gains.

Table I

F-TRANSFORM OF TYPICAL LOOP GAIN $T(s)$.

| Case | $T(s)$ | $\mathcal{F}[T(s)]$ (note: $p = \omega_p/\omega_s$ and $z = \omega_z/\omega_s$) |
|---|---|---|
| $\mathcal{C}_1$ | $\frac{1}{s+\omega_p}$ | $\frac{1}{\omega_s}\alpha(D,p) = \frac{1}{\omega_s}(\alpha_0(D) - \alpha_1(D)p + c(D,p))$ |
| $\mathcal{C}_2$ | $\frac{1}{s}$ | $\frac{1}{\omega_s}\alpha_0(D)$ |
| $\mathcal{C}_3$ | $\frac{1}{1+s/\omega_p}$ | $p\alpha(D,p)$ |
| $\mathcal{C}_4$ | $\frac{1+s/\omega_z}{1+s/\omega_p}$ | $p(1-\frac{p}{z})\alpha(D,p)$ |
| $\mathcal{C}_5$ | $\frac{1}{s(1+s/\omega_p)}$ | $\frac{1}{\omega_s}(\alpha_1(D)p - c(D,p)) = \frac{1}{\omega_s}(\alpha_0(D) - \alpha(D,p))$ |
| $\mathcal{C}_6$ | $\frac{1}{s^2}$ | $\frac{1}{\omega_s^2}\alpha_1(D)$ |
| $\mathcal{C}_7$ | $\frac{1+s/\omega_z}{s^2}$ | $\frac{1}{\omega_s^2}(\frac{1}{z}\alpha_0(D) + \alpha_1(D))$ |
| $\mathcal{C}_8$ | $\frac{1+s/\omega_z}{s(1+s/\omega_p)}$ | $\frac{1}{\omega_s}(\frac{p}{z}\alpha_0(D) - (\frac{p}{z}-1)(\alpha_1(D)p - c(D,p)))$ |
| $\mathcal{C}_9$ | $\frac{1+s/\omega_z}{s^2(1+s/\omega_p)}$ | $\frac{1}{\omega_s^2}(\frac{p}{z}\alpha_1(D) + (\frac{1}{p}-\frac{1}{z})c(D,p))$ |

From (10),

$$\omega_s \mathcal{F}[\frac{1}{s+\omega_p}] = -2\mathbf{Re}\left[\sum_{k=1}^{\infty}\frac{e^{j2k\pi D}}{jk+p}\right]$$
$$= 1/p - \pi e^{\pi p(1-2D)}\mathrm{csch}(\pi p) \quad (11)$$
$$:= \alpha(D,p)$$

where the proof of (11) can be obtained by looking up from a handbook of mathematical formulas or checked by a simple computer program, and csch is a hyperbolic function. Using Taylor series expansion, let $\alpha(D,p) = \sum_{k=0}^{\infty}(-1)^k\alpha_k(D)p^k$. Using the L'Hospital's rule, one has $\alpha_0(D) = \pi(2D-1)$ and $\alpha_1(D) = \pi^2(2D^2 - 2D + 1/3)$. A plot of $\alpha(D,p)$ is shown in Fig. 7. The right straight-line edge in Fig. 7 is $\alpha(D,0) = \alpha_0(D) = \pi(2D-1)$.

Let the correction term be $c(D,p) = \sum_{k=2}^{\infty}(-1)^k\alpha_k(D)p^k$. One has

$$c(D,p) = \alpha(D,p) - \alpha_0(D) + \alpha_1(D)p \quad (12)$$

A plot of $c(D,p)$ is shown in Fig. 8. From Fig. 8, the correction term $c(D,p)$ is significant only if $p > 0.2$ (equivalently, $\omega_p > 0.2\omega_s$). For $p < 0.2$, $c(D,p)$ can be ignored.

The F-transforms of other loop gains are shown in Table I. For each case $\mathcal{C}_1$-$\mathcal{C}_9$, the F-transform can be derived or proved (by simple algebra) in three ways. First, follow the definition of the F-transform as in (10). Second, by decomposing $T(s)$ into a combination of fractions $1/s$, $1/(s+\omega_p)$, etc, $\mathcal{F}[T(s)]$ is a combination of $\mathcal{F}[1/s]$, $\mathcal{F}[1/(s+\omega_p)]$, etc. For example,

- Use of $\mathcal{C}_1$ and $\mathcal{C}_2$ leads to $\mathcal{C}_5$ and $\mathcal{C}_8$;
- Use of $\mathcal{C}_2$ and $\mathcal{C}_6$ leads to $\mathcal{C}_7$;
- Use of $\mathcal{C}_1$, $\mathcal{C}_2$ and $\mathcal{C}_6$ leads to $\mathcal{C}_9$; and
- Use of $\mathcal{C}_1$ and the fact that $\mathcal{F}[1] = 0$ leads to $\mathcal{C}_4$.

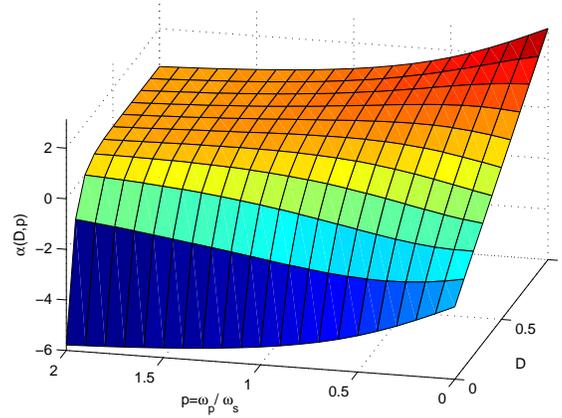

Figure 7. Plot of $\alpha(D,p)$ for case $\mathcal{C}_1$.

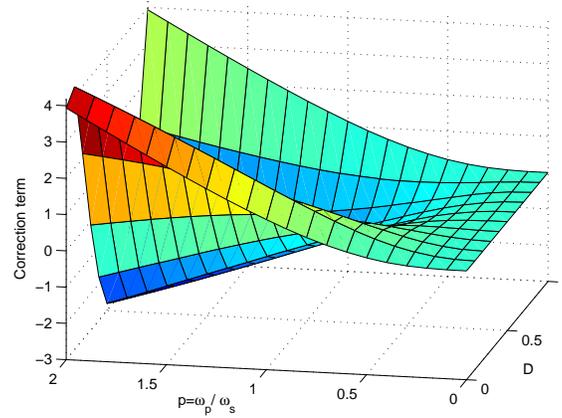

Figure 8. Plot of the correction term $c(D,p)$.

Third, each case is a special/general case of other cases. For example, by setting $\omega_p \to 0$, $\mathcal{C}_1$ leads to $\mathcal{C}_2$; $\mathcal{C}_5$ leads to $\mathcal{C}_6$; and $\mathcal{C}_8$ leads to $\mathcal{C}_7$. By setting $\omega_z \to \infty$, $\mathcal{C}_7$ leads to $\mathcal{C}_6$. One sees that $\mathcal{C}_1$ is a building block for other cases because they have *similar terms* as $\mathcal{C}_1$.

*Remarks:*

(a) All of the transforms in Table I are *exact*. No approximation is assumed.

(b) There is no correction term $c(D,p)$ for $\mathcal{C}_2$, $\mathcal{C}_6$, or $\mathcal{C}_7$. All other cases have a correction term $c(D,p)$, which is small and can be ignored if $p < 0.2$ as discussed above.

The systematic procedure to derive the critical conditions for two control schemes is discussed next. For other control schemes, the procedure is similar. Without loss of generality, assume ESR $R_c = 0$, then $\rho = 1$.

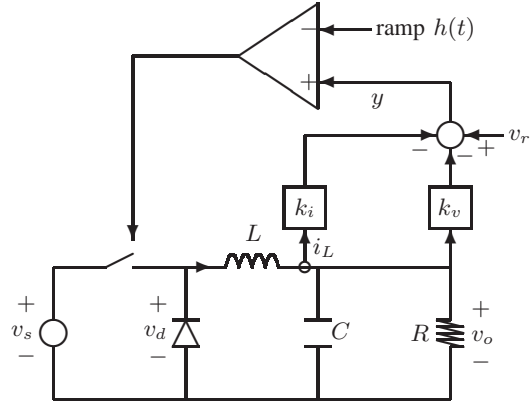

Figure 9. A buck converter under multi-loop state feedback.

## V. MULTI-LOOP STATE FEEDBACK: CASES $\mathcal{C}_2$ AND $\mathcal{C}_6$

Consider a buck converter under multi-loop state feedback, $y = v_r - k_i i_L - k_v v_o$, as shown in Fig. 9. From (1) and (2), one has $T(0) = k_i/R + k_v$ and

$$T(s) \approx \frac{v_s}{V_m}\left(\frac{k_i}{Ls} + \frac{k_v}{LCs^2}\right) \tag{13}$$

which is a combination of cases $\mathcal{C}_2$ and $\mathcal{C}_6$.

Let $K := 2L/RT$ which is a dimensionless parameter related to the loading as defined in [12]. From (9) and Table I, the critical condition is

$$\frac{v_s}{V_m}\left(\frac{k_i \alpha_0(D)}{L\omega_s} + \frac{k_v \alpha_1(D)}{LC\omega_s^2}\right) = \frac{v_s}{V_m}\left(\frac{k_i}{R} + k_v\right) + 1 \tag{14}$$

which can be arranged as

$$\frac{v_s k_i}{L}\left(D - \frac{K+1}{2}\right) + \frac{v_s k_v}{T}\left(-1 + \frac{T^2(1 - 6D + 6D^2)}{12LC}\right) = m_a \tag{15}$$

where the left side of (15) is also an approximate S-plot. For $T^2 \ll 12LC$, (15) becomes

$$\frac{v_s k_i}{L}\left(D - \frac{K+1}{2}\right) - \frac{v_s k_v}{T} = m_a \tag{16}$$

which can be rearranged as

$$D = \frac{K+1}{2} + \frac{Lm_a}{v_s k_i} + \frac{Lk_v}{Tk_i} \tag{17}$$

**Example.** Consider a buck converter under multi-loop state feedback [5, p. 232]. In [5], a digital control is used, but it can be approximated as a multi-loop analog control, with $y(t) = v_r - k_i i_L - k_v v_o$ intersecting with $h(t)$ to determine the switching actions. The converter parameters are $T = 400$ $\mu$s, $L = 20$ mH, $C = 47$ $\mu$F, $R = 22$ $\Omega$, $k_i = 2.1435$, $k_v = -0.1383$, $v_r = 0.2152$ V, and $V_m = 1$.

Using $v_s$ as the bifurcation parameter, the bifurcation diagram is shown in Fig. 10, and one sees that SNB occurs at $v_s = 20$ and $D = 0.7$. Generally in SNB, there is a hysteretic loop as shown in Fig. 10.

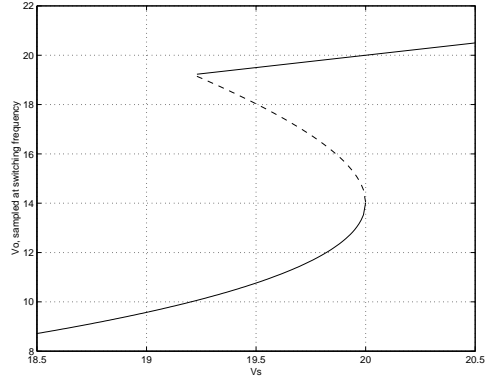

Figure 10. Bifurcation diagram showing stable (solid) and unstable (dashed) solutions. SNB occurs at $v_s = 20$, $D = 0.7$, and $v_o = v_s D = 14$.

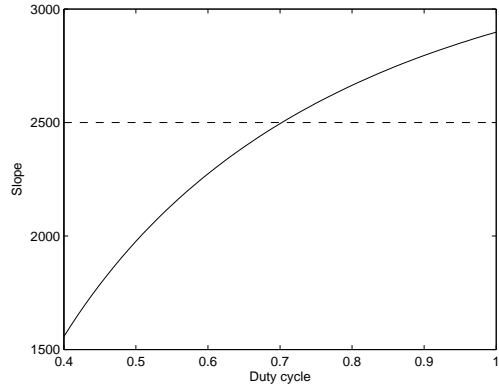

Figure 11. The S-plot (solid line) intersects with $m_a = 2500$ (dashed line) at $D = 0.7$ where SNB occurs.

In the figure, the upper solid line is for the operation when the switch is always on (hence $D = 1$), and the dashed line and the lower solid line are for unstable and stable $T$-periodic solutions respectively with a duty cycle less than 1. For $v_s$ between 19.25 V and 20 V, there are three solutions: a stable $T$-periodic solution, an unstable $T$-periodic solution, and the third (stable) solution for the switch being always on (hence $D = 1$). When the converter operates with the stable $T$-periodic solution and $v_s$ is increased a little above 20 V, the output voltage will *jump up* from 14 V to 20 V. Similarly, when the converter operates with $D = 1$ and $v_s$ is decreased a little below 19.25 V, the output voltage will *jump down* from 19.25 V to 10 V. The jumping up and down forms a hysteretic loop.

Using (7), the S-plot based on (15) (which is almost identical to the exact S-plot based on (8)) is shown in Fig. 11, which indicates that SNB occurs exactly at $D = 0.7$. The S-plot also shows that if the ramp slope is higher than 2898, the converter is stabilized. Also, from (17) and (7), one also has $D = 0.7$ when SNB occurs. □

## VI. CMC: Case $\mathcal{C}_2$

In CMC shown in Fig. 3, $y = i_c - i_L$. CMC can be considered as a special case of multi-loop control with $k_v = 0$ and $k_i = 1$. One has $T(s) \approx v_s/V_m L s$, which is of case $\mathcal{C}_2$. From (17) with $k_v = 0$ and $k_i = 1$, the critical condition is

$$D = \frac{K+1}{2} + \frac{Lm_a}{v_s} \tag{18}$$

It agrees with [13] which derives the same condition based on *either* sampled-data analysis or steady-state analysis. It is interesting to note that three (sampled-data, steady-state and harmonic balance) completely different analyses lead to the *same* condition. This also corroborates the accuracy of the derived critical condition (18).

Without the compensation ramp ($m_a = 0$), the critical point is $D = (K+1)/2$.

## VII. Conclusion

A general and exact SNB critical condition (9) in terms of the loop gain is derived. The critical condition is applicable to a *general* nonlinear switching system represented in Fig. 5. Therefore, it is also applicable to other similar nonlinear control systems. The effects of different parameters (such as $v_s$, $R$, and the ramp slope $m_a$) on the instability can be clearly seen. The derived critical conditions agreed with the past research results by sampled-data analysis or by time-domain simulations. The critical conditions also show the required ramp slope to stabilize the converter. The critical conditions have many different forms and lead to different plots. In the L-plot, the critical condition is $\mathcal{L} = \mathcal{F}[T(s)] = 1$. In the S-plot, the critical condition is $\mathcal{S} = m_a$. The derived critical conditions are helpful for the converter designers to predict or prevent some jump instabilities or coexistence of multiple solutions associated with the saddle-node bifurcation.

The questions asked in the Introduction are answered:

1) The previously known critical conditions, such as (18), become special cases in this generalized framework.
2) A typical critical condition is a weighted combination of three terms: $\alpha_0(D) = \pi(2D-1)$, $\alpha_1(D) = \pi^2(2D^2 - 2D + 1/3)$, and a correction term $c(D, p)$. Different control schemes have similar form of critical condition if the loop gain function is of the same case in Table I.
3) Given an *arbitrary* control scheme, a *systematic* procedure is proposed to derive the critical condition for that control scheme. First, approximate the loop gain $T(s)$ (for frequency higher than $\omega_s$). Then, from Table I, one can readily obtain the critical condition in terms of converter parameters.

Given the closed-form condition in terms of converter parameters, one knows the quantitative effect of each parameter on the instability. One can make the S-plot (such as Fig. 11) as a function of a parameter of interest, its intersection with the ramp slope $m_a$ determines the stable operating range of that parameter. The S-plot also shows the required ramp slope to stabilize the converter.

This paper focuses on the converter operated at a fixed switching frequency. Similar analysis can be applied to the converter with variable frequency control, such as constant on-time control. This paper also focuses on the SNB. Similar analysis can be also applied for PDB (subharmonic oscillation). The F-transforms in Table I still apply, but with a different $\alpha(D, p)$ and a minor modification. The results are reported separately.